\documentclass[showpacs,twocolumn,amsmath,amssymb,prb]{revtex4-1}

\usepackage[dvips]{graphicx}
\usepackage{wrapfig,subfigure}
\usepackage{amsmath,amsfonts,amssymb,amsthm}
\usepackage{fancyhdr}
\usepackage{mathrsfs}

\graphicspath{../Pictures/,../Pictures/Diagram/}

\begin{document}

\title{Lasing in circuit quantum electrodynamics with strong noise}

\author{ M. Marthaler$^1$,  Y. Utsumi$^2$ , D. S. Golubev$^{3,4}$}

\affiliation{$^1$Institut f\"ur Theoretische Festk\"orperphysik, Karlsruhe Institute for Technology, D-76128 Karlsruhe, Germany\\
               $^2$Department of Physics Engineering, Faculty of Engineering, Mie University, Japan\\
              $^3$Institut f\"ur Nanotechnologie, Karlsruhe Institute of Technology,D-76021 Karlsruhe, Germany \\
             $^4$O.V. Lounasmaa Laboratory, Aalto University School of Science, Finnland}

\begin{abstract}
 We study a model which can describe a superconducting single electron transistor (SSET) or a double quantum dot coupled to transmission-line oscillator.
 In both cases the degree of freedom is given by a charged particle, which couples strongly to the electromagnetic environment or phonons. We consider the
 case where a lasing condition is established and study the dependence of the average photon number in the resonator on the spectral function
 of the electromagnetic environment. We focus on three important cases: a strongly coupled environment with a small cut-off frequency, 
 a structured environment peaked at a specific frequency and 1/f-noise.  We find that the electromagnetic environment can have a substantial impact on the 
 photon creation. Resonance peaks are in general broadened and additional resonances can appear. 
\end{abstract}

\maketitle

\section{Introduction}

\begin{figure}[t]
\includegraphics[width=6 cm]{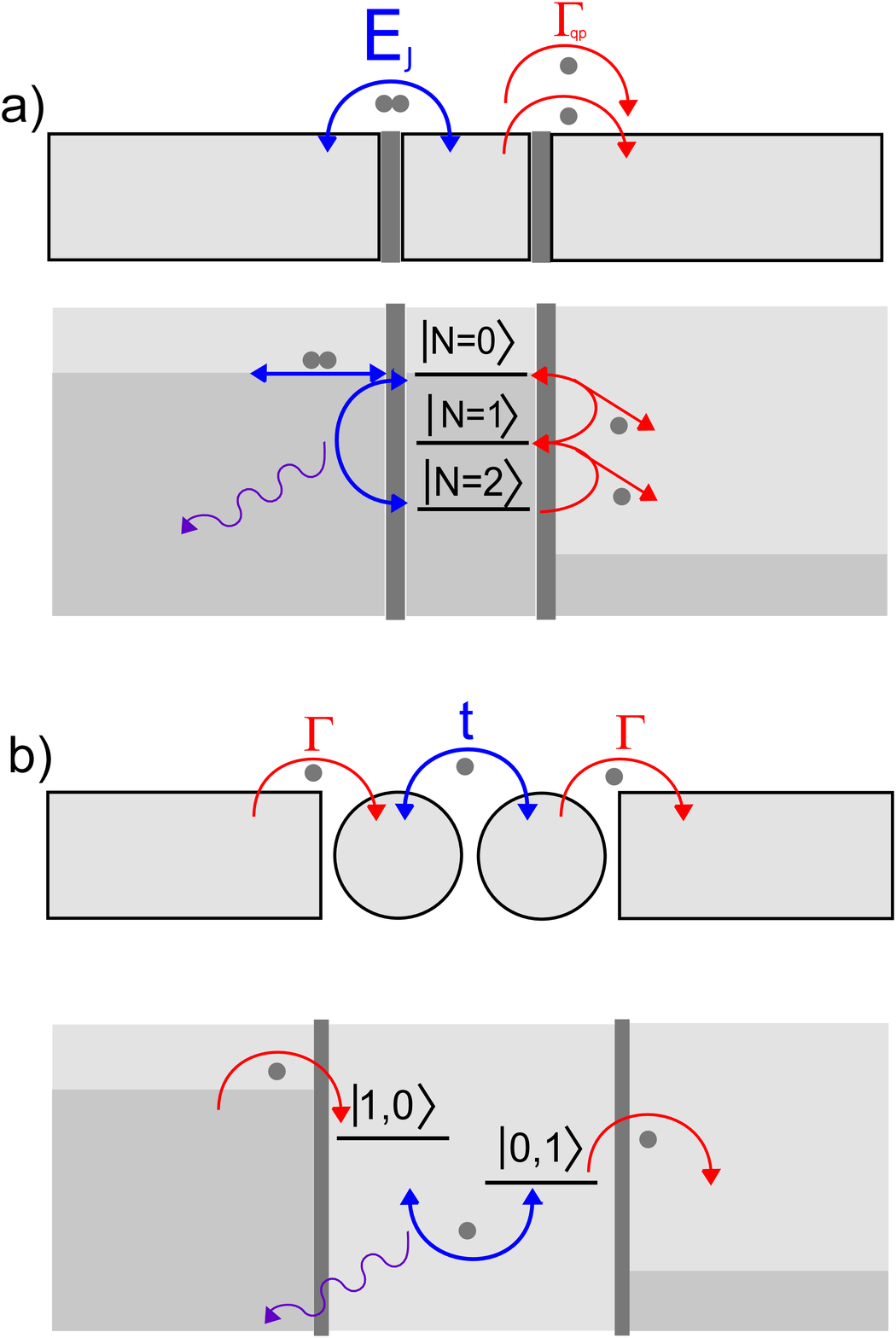}
\caption{A schematic depiction of the lasing cycle for the SSET and for the double dot. 
(a) A cooper pair tunnels trough the left junction onto the island and via
 two consecutive single charge tunneling events into the right lead. 
The energy for the process is provided by the transport voltage $V$, 
and the gate voltage $V_G$ can be used to change the gate charge $\delta N_{\rm G}$. 
The energy of the island with two additional charges is tuned to be smaller than the 
state with zero additional charges. The energy difference, is approximately in resonance with a cavity coupled to the island. 
(b) An electron tunnels from the left lead into the left dot. Then the electron tunnels from left to right dot, emits a photon and
finally tunnels into the right lead.
}
\label{fig:cycle}
\end{figure}

 In recent experiments \cite{Astafiev2007,JJ_Laser_second_realization_Armour} 
 an effective single-atom maser was realized
 using a superconducting single electron transistor (SSET),
 which is essentially a charge qubit with an additional 
 applied transport voltage.
 The two qubit states are coupled to a transmission-line resonator and a 
 third state plays a role in the pumping cycle [Fig. \ref{fig:cycle} (a)]. This
 system has been theoretically studied \cite{Rodrigues2007,Blencowe2005,Armour2005,Clerk2005,Brosco2009}  and
 it has been experimentally shown \cite{Astafiev2007,JJ_Laser_second_realization_Armour}
 that lasing can be achieved. Theoretically it has also been shown that 
 a SSET can be used to create a non-classical photon distribution in a resonator \cite{SSET_Nonclassical_marthaler,Nonclassical_marthaler_2}. 
 The same design can be used with
 a mechanical oscillators \cite{Naik2006, Schwab2010} instead of 
 a transmission-line resonator and has been considered as one of the possibilities 
 to observe non-classical states
 in such a macroscopic object \cite{Blencowe2004}. If a Josephson 
 junction is directly shunted by a resonator it is also possible to observe enhanced photon creation \cite{Hofheinz_JJosci},
 multistability \cite{Armour_JJosci}, and non-classical effects \cite{Juha_JJosci_1,Juha_JJosci_2,Ankerhold_JJosci}.

 A very similar situation has been studied using a double quantum dot coupled
 to a transmission-line resonator \cite{Ensslin_Osci_DDot_v1,Ensslin_Osci_DDot_v2} [Fig. \ref{fig:cycle} (b)]. 
 Here an applied transport voltage drives a current through a double dot, where the energy difference between left and right dot can be used to 
 create photons. 
 The correlation of photon emission 
 and transport properties has been extensively studied \cite{Marthaler_Jin_DDotLasing,Bergfeld2013,FlindtLambert2013,Kontos_2014}.
 It has also been theoretically shown
 that lasing can be achieved \cite{Marthaler_Jin_DDotLasing,Marthaler_Jinshuang_DDotLasing}.
 Already, in experiments a linewidth
 narrowing has been observed \cite{Peta_Lasing_DDot} and even clear experimental evidence for lasing has been shown \cite{Peta_Lasing_Science}.
 Many theoretical studies have been performed for this setup.
 It has e.g. been used as a toy model for the quantum photovoltaic effect \cite{Vavilov_2013}
 and it has been considered as a hybrid microwave-cavity heat engine \cite{Bergenfeldt_Hybvrid_Heat}.

 In both cases, SSET and  double dot, we consider the charge degree of freedom. Charge is strongly coupled
 to any perturbation from the environment. The charge variable of the SSET is strongly coupled to 
 fluctuations in the background charge
 and the electromagnetic environment \cite{Pekola2009,Frey2014_chargenoise}. In the double dot, the electron is strongly coupled
 to phonons \cite{Ensslin_Phonons}.
 In most theoretical studies the effects of noise are modeled by using a lowest order master equation, which assumes a smooth spectral function
 and relatively weak coupling to the environment. However, it has already been shown that if strong coupling to the bath
 and higher order transition processes are considered, 
 new effects like enhanced photon emission \cite{Peta_Theory_enviornment_enhanced_emission} 
 and inversionless lasing \cite{Marthaler_Inversionless} can appear.

 We are going to consider strong coupling to the environment with a structured spectral function.
 The approach used in this paper is similar to the description of strong dephasing in 
 a quantum dot coupled to a cavity \cite{Wilson-Rael2002}. However
 we consider a system where noise creates both dephasing and relaxation,
 depending on the bias point. Additionally we consider charge tunneling (either quasiparticles for the SSET of electrons for the double dot),
 which serves as pumping mechanism for the three level maser. We will
 focus on three specific examples of noise spectral densities:
 low frequency noise created by a strong 
 external resistor \cite{Dutta1981}, finite frequency resonances \cite{SpuriousResonances,Lindell2003} 
 and 1/f noise  \cite{Mueck2005,NakamuraTsai_onef2,Altshuler_onef3}. 
  Our results suggest that low frequency noise
 was an important factor in the realization of the single artificial atom maser  using an SSET \cite{Astafiev2007}
 and that spurious resonances
  can lead to additional resonance peaks.
 In certain limiting cases it is possible to describe the 
 system by a Markovian master equation, that treats  photon creation as stochastic
 process. Despite the fact that in this case all quantum coherence is
 lost, the classical coherence of radiation, which is the hallmark of lasing is still preserved and we will explicitly show the connection
 of our approach to a standard description of the laser \cite{Scully} in section \ref{sec_one_over_f_noise}.

  In the next section we will explain in detail the model we use to describe the SSET coupled to the resonator and to strong charge noise.
  We write the Hamiltonian
 in the basis of the charge qubit eigenstates
 and use a standard description of a bosonic reservoir, coupled to the island charge. 
 Then we discuss the consequences of strong charge noise and introduce the diagonalisation
 of the longitudinal charge noise component. Noise creates transitions between the states of our system.
 We will discuss these rates and introduce a master equation in the Lindblad form which 
 governs the dynamics of our system. In a short section we will than show how exactly the same model can be used to describe
 a double dot coupled to a resonator. 
 The next two sections will be used to 
 describe the effects of coupling to a strong external resistor, to a single mode in equilibrium
 and to $1/f$ noise. In the end we summarize our results and we shortly discuss their consequences 
 in regard to experiments on mechanical oscillators coupled to single charge devices and for further work on lasing.

\section{Model Hamiltonian}

 In this section we will first discuss the Hamiltonian of a 
 SSET in great detail. 
 The SSET consists of two superconducting leads
 coupled by tunnel junctions to a superconducting island.
 A gate voltage $V_G$ shifts the
 electrostatic energy of the island and controls,
 together with the transport voltage $V$, the current through the device.
 The Josephson coupling $E_{\rm J}$ of the junctions should be weak compared
 to the charging energy scale, $E_C= e^2/2C_{\Sigma}$ ($C_{\Sigma}$ is the total
 capacitance of the island),
 and the superconducting energy gap $\Delta$. It leads
 to coherent Cooper pair tunneling, with pronounced consequences when two
 charge states differing by one Cooper pair are nearly degenerate.
 In addition, quasiparticles tunnel incoherently (with rate
 $\propto V/eR$, where $R$ is the normal state resistance on the junction)  when
 the energy difference between initial and final state
 is sufficient to create a quasiparticle excitation, i.e., when it
 exceeds twice the gap
 (assumed equal for electrodes and island), $|\Delta E |\ge2\Delta$ \cite{Maassen1991}.
 The gate charge $N_G=C_G V_G/e$ is chosen to be close to one, $\delta N_G=N_G-1< 1/2$.
 This means that we tune the SSET close to the JQP cycle
 where current is transported in the following way:
 A Cooper-Pair tunnels through the left junction onto the island, increasing the
  island charge $N$
 by $2$. This is 
 followed by two quasiparticle tunneling events 
 through the right junction, that brings the island back to its original
 configuration \cite{Choi2003} [Fig. \ref{fig:cycle} (a)].

To create a laser-like situation in the SSET
 we utilize two charge states
 $|N=0\rangle$ and $|N=2\rangle$. These two states
 are coupled by Josephson tunneling through the left junction. 
The similarities to quantum optical systems becomes more obvious in the base of the charge
 qubit, where we diagonalize the coupled charge states. 
Additionally it is necessary to create population inversion. 
This is achieved by using a third state, the odd charge state $|N = 1 \rangle$.

We describe the system by the following Hamiltonian, 
\begin{eqnarray}
H &=& H_{\rm sys} + H_{\rm env} + H_{\rm qp} + H_{\rm diss}
\nonumber \\
& & +V_{\rm coupl}  + V_{\rm qp} + V_{\rm diss}.
\label{H}
\end{eqnarray}
The system is given by a
qubit and a single mode oscillator in the electromagnetic resonator,
{\nolinebreak $H_{\rm sys}=H_{\rm qbit}+H_{\rm osc}$.}
The first contribution, $H_{\rm qbit}$, describes an isolated qubit 
in the form
\begin{eqnarray}
H_{\rm qbit}
&=&
\sum_{N=0}^2
E_C (N_G-N)^2 P_{N,N}
\nonumber \\ & &
-
\frac{E_J}{2} 
(P_{2,0}+P_{0,2})
,
\end{eqnarray}
where we introduced projection operators, 
\begin{eqnarray}
P_{N,M} = |N \rangle \langle M| 
\end{eqnarray}
and $|N\rangle$ are the charge states of the island.

\begin{figure}
\includegraphics[width= 0.8 \columnwidth]{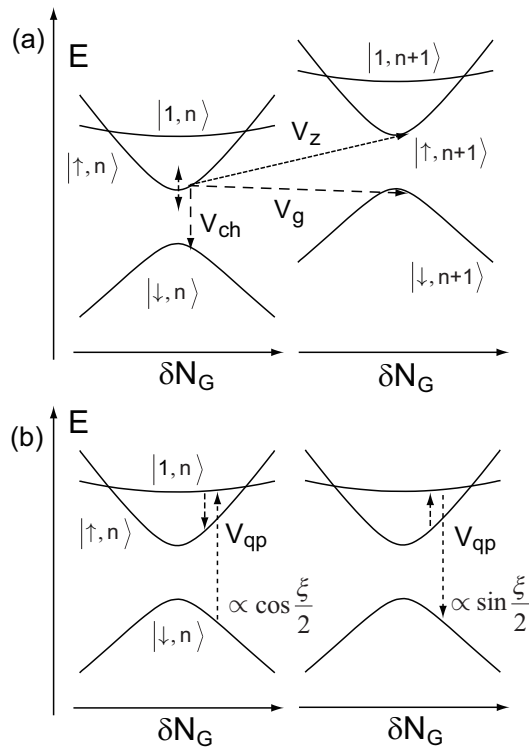}
\caption{
(a) 
Energy diagram for the SSET and the electromagnetic
 resonator with $n$ (left panel) and $n+1$ photons (right panel) at the resonance condition 
$\delta \omega=\omega_0-\Delta E = 0$. 
Two states 
$|\! \uparrow,n \rangle$ and 
$|\! \downarrow,n+1 \rangle$
are energetically degenerate. 
The odd charge states 
$|1,n \rangle$ and 
$|1,n+1 \rangle$ are  shown at the energy $-E_C+eV$.
The longitudinal coupling to the bosonic bath $V_{\rm z}$, 
which causes dephasing, 
increases away from the symmetry point 
$\delta N_{\rm G}=0$. 
It may spoil the transverse coupling to the oscillator $V_{\rm g}$ crucial for coherent lasing. 
The impact of the longitudinal coupling to the bosonic bath $V_{\rm ch}$, 
which causes relaxation, 
will be small since the states
$|\! \downarrow,n \rangle$ and $|\! \uparrow,n \rangle$ are
energetically separated by the gap $\Delta E$. 
(b)
Creation of population inversion by quasiparticle tunneling. 
For $\delta N_G >0$, the condition
$\cos (\xi/2) > \sin (\xi/2)$
is satisfied. 
The probability for the pumping up process 
$|\! \downarrow,n \rangle \to |\! 1,n \rangle \to |\! \uparrow,n \rangle$ 
(left panel) 
is higher than that for the pumping down process
$|\! \uparrow,n \rangle \to |\! 1,n \rangle \to |\! \downarrow,n \rangle$
(right panel). 
At the symmetry point, 
 $\delta N_G =0$, 
both processes occur equally,
$\cos (\xi/2) = \sin (\xi/2)=1/\sqrt{2}$,
and thus no population inversion is created. 
}
\label{fig:diagram}
\end{figure}

 The two charge state $N=0$ and $N=2$ are coupled by Josephson tunneling through the left junction
 with strength $E_J$ [Fig. \ref{fig:cycle} (a)].
 Across the right junction we apply a large voltage drop, therefore the Josephson 
 tunneling for this Junction
 can be neglected \cite{Choi2003}. 
 We diagonalize the qubit part of the Hamiltonian $H_{\rm qbit}$ by introducing the states,
\begin{eqnarray}\label{eq_Upanddownarrow_total_states_at_beginning}
|\! \uparrow \rangle   &=& \cos\frac{\xi}{2}|N=0\rangle + \sin\frac{\xi}{2}|N=2\rangle\, ,\\
|\! \downarrow \rangle &=& \sin\frac{\xi}{2}|N=0\rangle - \cos\frac{\xi}{2}|N=2\rangle\, ,\nonumber\\
\tan\xi &=& 
\frac{E_{\rm J}}{4 \, \delta N_{\rm G} E_C} \, .\nonumber
\end{eqnarray}
The states $|\!\uparrow\rangle$ and 
$|\!\downarrow\rangle$ play the essential role to generate lasing
and 
$|N=1\rangle$ is needed to create population inversion. 
It is convenient to introduce Pauli matrices, 
\begin{eqnarray}
\sigma_z 
&=& 
|\! \uparrow \rangle \langle \uparrow \!| -
|\! \downarrow \rangle \langle \downarrow \!|,
\\
\sigma_x 
&=& 
\sigma_+ + \sigma_-, \nonumber
\\
\sigma_+
&=&
\sigma_-^\dagger
=
|\! \uparrow \rangle \langle \downarrow \!| \, . \nonumber
\end{eqnarray}
 and 
the diagonal Hamiltonian of the qubit in this basis; 
\begin{eqnarray}\label{eq_SSET_in_the_form_of_a_twolevelsystem}
H_{\rm qbit}
&=&
\frac{
\Delta E \sigma_z
}{2} 
-E_C
\, P_{1,1}
\, ,
\\
\Delta E &=& 
\sqrt{E_J^2+4(E_C\delta N_G)^2}, \nonumber
\end{eqnarray}
where we shifted the origin of energy.
The Hamiltonian of the electromagnetic resonator is 
\begin{eqnarray}
H_{\rm osc} = \omega_0 \, a^\dagger a \, .
\end{eqnarray}
The eigenstate of the system Hamiltonian
$H_{\rm sys}$ 
is characterized by the qubit states, $\sigma=\uparrow,\downarrow,1$, 
and the number of photons $n$ as $| \sigma, n \rangle$
with the energy 
$E_\sigma+n \, \omega_0$ 
where 
$E_{\uparrow/\downarrow}=\pm \Delta E/2$
and 
$E_1=-E_C$. 
For lasing, the qubit is operated 
around the resonant condition, 
\begin{eqnarray}
\delta \omega = 0 ,
\;\;\;\;
\delta \omega \equiv \omega_0 - \Delta E,
\end{eqnarray}
so that the states 
$|\! \downarrow, n+1 \rangle$ 
and 
$|\! \uparrow, n \rangle$ 
are degenerate 
[Fig.~\ref{fig:diagram} (a)]
and thus coherent emission or absorption of photons can be induced.

For photon emission, population inversion 
has to be created by quasiparticle tunneling \cite{Rodrigues2007,Blencowe2005}. 
The Hamiltonian of the quasiparticles reads
\begin{eqnarray}
H_{\rm qp}=\sum_{r=R,I}\sum_i \epsilon_{i,r}c_{i,r}^\dagger c_{i,r}, 
\end{eqnarray}
 where the subscript $R$ denotes the right lead,
 while the subscript $I$ denotes the superconducting island. 
 We only apply a large voltage drop across the right junction, therefore we neglect
 quasiparticle tunneling through the left junction \cite{Choi2003}.
 The effect of the tunneling of quasiparticles on the qubit is described by the following Hamiltonian
\begin{eqnarray}
V_{\rm qp}
&=&
T_{\rm qp} \, (P_{2,1}+ P_{1,0}) + {\rm H.c.}
\, ,
\\
T_{\rm qp} &=& \sum_{i,k} t^{\rm qp} \, c_{i,I}^\dagger c_{k,R}.\nonumber
\end{eqnarray}
The quasiparticles obey the Fermi distribution 
\begin{eqnarray}
f^+(\omega)
=
\frac{1}{{\rm e}^{\beta_e \omega}+1} \, .
\end{eqnarray}
with inverse electron temperature $\beta_e$.
The density of states of quasiparticles is gapped, 
\begin{eqnarray}
\rho(\omega)
=
\sum_{i} \delta(\omega-\epsilon_{i,r})
=
\rho_0
\frac{
\theta(1-\Delta^2/\omega^2)
}
{\sqrt{1-\Delta^2/\omega^2}}
\, ,
\end{eqnarray}
and thus only for $eV > 2 \Delta$ quasiparticle tunneling and simultaneous change in the charge states occurs. 
In the qubit basis, it can be rewritten as, 
\begin{eqnarray}
V_{\rm qp}
&=&
T_{\rm qp} \, 
\left[
\cos \frac{\xi}{2} (-P_{\uparrow,1}+P_{1,\downarrow})+
\right.
\nonumber \\
& &
\left. 
+\sin \frac{\xi}{2} (P_{\downarrow,1}+P_{1,\uparrow})
\right] + {\rm H.c.}
\, .
\end{eqnarray}
For $\delta N_G >0$, the condition 
$\cos (\xi/2) > \sin (\xi/2)$ 
is satisfied and 
the probability for the pumping process 
$|\! \downarrow,n \rangle \to |1,n\rangle \to |\! \uparrow,n \rangle$ 
[Fig.~\ref{fig:diagram}(b) left panel] 
is higher than that for the anti-pumping process
$|\!\!\uparrow,n \rangle \to |1,n \rangle \to |\!\!\downarrow,n \rangle$
[Fig.~\ref{fig:diagram}(b) right panel]. 
At the symmetry point, $\delta N_G =0$, 
both process occur equally
$\cos (\xi/2) = \sin (\xi/2)=1/\sqrt{2}$
and thus no population inversion is created.

The qubit is capacitively coupled with the transmission-line oscillator. 
It is also coupled with environmental charge fluctuations. 
These couplings appear in the same form as, 
\begin{eqnarray}\label{eq_V_coupl_to _charge_noise}
V_{\rm coupl} 
&=&
[g \, (a^\dagger +a) + x] \, (P_{0,0}-P_{2,2}),
\\
x
&=&
\sum_i \frac{t_i}{2} (b^\dagger_i+b_i) ,\nonumber
\end{eqnarray}
where we introduce another oscillator bath to model the environmental charge fluctuations, 
\begin{eqnarray}
H_{\rm env} &=& \sum_i \omega_i b_i^\dagger b_i.
\end{eqnarray} 
The bosonic bath is characterized by the correlation function, 
\begin{eqnarray}
\langle x(t) x(0) \rangle
=
\frac{1}{4}
\int \frac{d \omega}{2 \pi}
N(\omega) \, n^-(\omega)
\, e^{-i \omega t}
\, .
\end{eqnarray}
The equilibrium distribution function for bosons reads, 
\begin{eqnarray}
n^-(\omega)=\frac{1}{1-{\rm e}^{-\beta \omega}} \, , 
\end{eqnarray}
where $\beta=1/k_B T$ and $T$ is the noise temperature. 
The spectral density, 
\begin{eqnarray}
N(\omega)=2 \pi \sum_i |t_i|^2 
[
\delta(\omega-\omega_i)
-
\delta(\omega+\omega_i)
] \, , 
\end{eqnarray}
can be arbitrary. 
This model is versatile and able to describe a variety of charge noise sources in equilibrium. 
For example, if the source of the environmental charge fluctuation 
is the external circuit 
the spectral function is given as, 
\begin{eqnarray}\label{eq_N_of_omega_as_function_of_Z}
N(\omega)=\omega {\rm Re} Z(\omega)/R_K
\, . 
\end{eqnarray}
where $Z(\omega)$ is the effective impedance as seen from the quantum system
\cite{grabertpofe,SingleChargeTunneling}, and $R_K$ is the resistance quantum. 
The electromagnetic environment is known to have a significant 
impact for single charge devices \cite{Nazarov1989}. 
 This model  is also able to describe the coupling to another
 mode in the electromagnetic resonator with frequency $\omega_L$,
 which is off-resonant and thus in equilibrium. 
 The spectral density is, 
\begin{eqnarray}
N(\omega)=\omega_L \epsilon_C 
[
\delta(\omega-\omega_L)
-
\delta(\omega+\omega_L)
] \, , 
\end{eqnarray}
 where $\epsilon_C$ characterizes the coupling strength.
 For an extensive discussion of the form of the effective
 impedance and its relation to the coupling to the electromagnetic environment see Ref.~[\onlinecite{grabertpofe}].

In the basis of the diagonal qubit Hamiltonian, the couplings between 
the qubit and the oscillator consists of a longitudinal and a transverse part,
$V_{\rm coupl} = V_z+V_\perp$. 
The longitudinal coupling reads, 
\begin{eqnarray}
V_z &=& [g \, (a^\dagger +a) + x] \,  \cos \xi \sigma_z , 
\end{eqnarray}
and we separate the transverse coupling into two parts 
$V_\perp = V_{\rm ch}+V_{\rm g}$ 
with 
\begin{eqnarray}
V_{\rm ch} &=& x \,  \sin \xi \sigma_x , 
\\
V_{\rm g} & \approx & 
g (a^\dagger \sigma_- + a \sigma_+) \sin \xi  , \nonumber
\end{eqnarray}
where we performed the rotating wave approximation, namely we neglect terms 
$a^\dagger \sigma_+$
and 
$a \sigma_-$ 
connecting the states 
$|\! \uparrow,n \rangle$ and 
$|\! \downarrow,n-1 \rangle$ 
energetically separated by $\Delta E+\omega_0$. 
Now $V_{\rm g}$ induces transitions between the qubit states 
$|\! \uparrow,n \rangle$ and 
$|\! \downarrow,n+1 \rangle$ 
[Fig.~\ref{fig:diagram} (a)]. 
The transverse coupling to the bosonic bath $V_{\rm ch}$ causes relaxation 
[Fig.~\ref{fig:diagram} (a)],
while the longitudinal coupling $V_{\rm z}$ 
cause dephasing. 
The relaxation effect $V_{\rm ch}$ is weak
since the states $|\! \uparrow,n \rangle$ and $|\! \downarrow,n \rangle$ 
are energetically separated by the gap $\Delta E$ and the relevant
frequencies of the spectral densities considered in this paper are all smaller than $\Delta E$.
Further by tuning to the symmetry point $\delta N_{G}=0$ ($\xi=\pi/2$), 
the longitudinal coupling $V_{\rm z}$ disappears.
However, for lasing, in order to create population inversion, 
the qubit has to be tuned away from symmetry, 
$\delta N_{\rm G} > 0$. 

Finally, we also account for the mixing of the single mode in the resonator to an external circuit again modeled by oscillators, 
\begin{eqnarray}
H_{\rm diss} &=& \sum_i \omega_i^d d_i^\dagger d_i, 
\\
V_{\rm diss} &=& \sum_i t^d_i d^\dagger_i a + H.c. \, .\nonumber
\end{eqnarray}
Since the precise form of the spectrum is not relevant we simply assume it smooth around the frequency of the resonator, which leads to a decay rate of the form
\begin{eqnarray}
\kappa = \int \!\! dt \sum_i |t_i^d|^2 \langle d_i(t)d_i^{\dag}(0)\rangle 
\, e^{i\omega_0 t},
\end{eqnarray}
The temperature in the oscillator reservoir does not have to coincide with the
noise temperature and since it does not substantially effect our results we choose the temperature
in the oscillator to be small.

\subsection{Lasing-like behavior in incoherent photon emission}

In order to generate lasing, 
first the qubit is tuned to the resonant condition, 
where 
$|\!\! \uparrow,n \rangle$ and 
$|\!\! \downarrow,n+1 \rangle$
are energetically degenerate [Fig.~\ref{fig:diagram} (a)], 
and then photon emission or absorption is possible. 
This condition must be fulfilled at a bias point $\delta N_G >0$, 
such that population inversion is created. 
At the same time, it inevitably creates longitudinal
 coupling between the qubit and charge noise  $V_z$. 
The size of the fluctuations is characterized by 
\begin{eqnarray}
\cos^2 \! \xi \, \langle x(t)^2 \rangle
=
\frac{\cos^2 \! \xi}{4}
\int \!\! \frac{d \omega}{2 \pi} \, N(\omega) n^-(\omega). 
\end{eqnarray}
Then the longitudinal coupling with the bosonic field $V_z$ 
competes with the transverse coupling with the oscillator $V_{\rm g}$. 
We consider the case when the fluctuations in the level
 splitting can be larger than the Rabi frequency, the transverse coupling between oscillator and SSET, 
\begin{equation}
g \sqrt{\langle n\rangle} \sin\xi
<
\sqrt{\langle x(t)^2 \rangle} \cos\xi \, ,
\end{equation}
where $\langle n \rangle=\langle a^{\dag} a \rangle$ is the average number of photons in the system. 
As we tune away from degeneracy the transverse coupling to the oscillator
 that is needed to create photons, decays as $\sin \xi$ at the same time, 
 the fluctuations in the level splitting grows as $\cos \xi$. 
We diagonalize the longitudinal part $V_z$, 
using a polaron transformation~\cite{Mahan,Weiss}, 
\begin{eqnarray}
U=
\exp \left[ -i \cos \xi (p+p')  \, \sigma_z \right]
. 
\end{eqnarray}
Here, in order to write the resulting Hamiltonian in a compact form we introduce
the operators, 
\begin{eqnarray}
p = i \sum_i \frac{t_i \, (b_i^{\dag}-b_i)}{2 \, \omega_i}
\, , 
\;\;\;\;
p' = i \, \frac{g \, (a^{\dag}-a)}{\omega_0}
\, .
\end{eqnarray}
After the transformation the coupling between SSET the oscillator and charge noise takes the form
\begin{eqnarray}
\tilde{V}_{\rm coupl} 
=
U {V}_{\rm coupl} U^\dagger
=
\tilde{V}_{\rm ch}
+
\tilde{V}_{\rm g}
,
\label{eq_H_couple_Polaron}
\end{eqnarray}
where 
$\tilde{V}_{\rm g}$ and 
$\tilde{V}_{\rm ch}$ 
are now dressed by polarons as, 
\begin{eqnarray}
\tilde{V}_{\rm g}
\! & \approx & \!
g \sin\xi 
e^{-i S} 
e^{-i S'/2} \, a \, e^{-i S'/2} \sigma_+
+{\rm H.c.} \, ,
\\
\tilde{V}_{\rm ch}
\! &=& \!
\sin\xi 
e^{-i S'} 
e^{-i S/2} \, \delta u \, e^{-i S/2} \sigma_+
+{\rm H.c.} \, ,\nonumber
\end{eqnarray}
where 
$S = p \cos \xi$
and 
$S' = p' \cos \xi$. 
Here $\tilde{V}_{\rm g}$ may be treated as the perturbation. 
$\tilde{V}_{\rm ch}$ may be also treated as  perturbation because 
the relevant frequencies of the spectral functions considered are smaller than the 
energy splitting $\Delta E$.  
The quasiparticle tunneling is dressed with polarons as well and the transformed Hamiltonian reads, 
\begin{eqnarray}
\tilde{V}_{\rm qp}
&=&
T_{\rm qp} \, 
\left[
\cos \frac{\xi}{2} 
\left(
-e^{-i (S+S')/2} P_{\uparrow,1}
\right.
\right.
\nonumber \\
& &
\left.
+P_{1,\downarrow} e^{-i (S+S')/2}
\right)
+
\sin \frac{\xi}{2} 
\left( 
e^{i (S+S')/2} P_{\downarrow,1}
\right. 
\nonumber \\
& &
\left. 
\left. 
+P_{1,\uparrow} e^{i (S+S')/2}
\right)
\right] + {\rm H.c.} \, 
\label{eq_Hqp_qbit_Full}
\end{eqnarray}
However, quasi-particle tunneling processes are dominated by the energy scale of the transport voltage $eV$
 and phase fluctuations cause only minor effects as we will discuss later.

Here at this point we summarize the non-perturbative part $H_0$ and 
the perturbative part $H'$, 
\begin{eqnarray}
H_0 &=& H_{\rm sys} + H_{\rm env} + H_{\rm qp} + H_{\rm diss} \, ,
\\
H' &=& \tilde{V}_{\rm g} +  \tilde{V}_{\rm ch} + 
\tilde{V}_{\rm qp} + V_{\rm diss} \, .\nonumber
\end{eqnarray}
Then the reduced density matrix is obtained by tracing over all reservoir 
degrees of freedom, consisting of the charge noise, quasiparticles and dissipation in the oscillator 
$H_{\rm env} + H_{\rm qp} + H_{\rm diss}$
; 
\begin{eqnarray}
\rho(t) &=& 
{\rm Tr}_{\rm R} 
\left[
{\rm e}^{-i (H_0+H') t} \rho_0 {\rm e}^{i (H_0+H') t}
\right] 
\, ,
\\
\rho_0 & \propto & {\rm e}^{-\beta H_0}
\, ,\nonumber
\end{eqnarray}
where the initial density matrix is normalized as
${\rm Tr} \rho_0=1$.

Under the situation we are considering, because of strong environmental noise, the coherent coupling between the states 
$| \! \uparrow, n \rangle$
and 
$| \! \downarrow, n+1 \rangle$
is lost and transitions among different photon states becomes stochastic. 
The quantities of interest are the average number of photons $\langle n \rangle =\langle a^{\dag} a\rangle$ and the Fano-Factor 
\begin{eqnarray}
F = \frac{\langle n^2\rangle-\langle n\rangle^2}{\langle n\rangle}\, .
\end{eqnarray}
The Fano-Factor defines the width of the photon number distribution. 
In a pumped oscillator one would expect a photon number distribution, 
similar to a Poisson distribution, in this case the Fano-Factor should be 
close to $F=1$. If we only heat the oscillator and we have a thermal photon 
number distribution the Fano-Factor should be close to $F=\langle n\rangle+1$.

\subsection{Transition rates}

 Let us take a closer look at few selected transition rates. We do this
 to make the connection between our microscopic
 Hamiltonian and our Markovian Master equation clear. 
 First, let us consider the quasiparticle tunneling processes for generating the population inversion. 
 The transition rate of the process 
 $| 1, n \rangle \to |\downarrow, n \rangle$
 within Fermi's golden rule in 
 $\tilde{V}_{\rm qp}$ reads, 
\begin{eqnarray}
\Gamma_{| \downarrow, n \rangle \leftarrow |1, n \rangle}
\!\! &=& \!\!
2 \cos^2 \frac{\xi}{2} 
\, {\rm Re} \!\!
\int_0^\infty \!\!\!\! dt \, 
\gamma_{\rm qp}^{+-}(t) \, 
C^n_{n,1/2}
\nonumber \\
& & \times
C_{\rm g}(t)^{1/4} \, 
{\rm e}^{i(E_{\downarrow,n}-E_{1,n}-eV)t}
\, .
\end{eqnarray}
where the particle-hole propagator is given by, 
\begin{eqnarray}
\gamma_{\rm qp}^{+-}(\omega)
\!\! &=& \!\!
\langle T_{\rm qp}(t) T_{\rm qp}^\dagger  \rangle 
=
2 \pi |t^{\rm qp}|^2
\!\! \int\, \int\!\! d \omega'd\omega
\rho(\omega')
\nonumber \\ 
& &
\times
\rho(\omega+\omega')
f^-(\omega+\omega')
f^+(\omega')e^{i(\omega+\omega')t}
,\nonumber\\
\end{eqnarray}
 where 
 $f^- \!=\! 1 \!-\! f^+$ and $f^+$ is the
 Fermi distribution. 
 The matrix element is given by
 $
 C^n_{n,\lambda}
 \!=\!
 |\langle n| {\rm e}^{-i \lambda S'} |n \rangle|^2\approx 1$
 up to the order of $(g/\omega_0)^2$. For all further purposes we will ignore the effect of higher orders of $(g/\omega_0)^2$
 on the quasiparticle tunneling rate, since it has no significant impact in our regime. For a discussion
 of effects this coupling to the oscillator can have see e.g. Ref. \onlinecite{FranckCondon}.
 The correlation function $C_{\rm g}(t)$
 is well known from $P(E)$-Theory~\cite{SingleChargeTunneling}; 
\begin{eqnarray}
\label{eq_gcorel}
C_{\rm g}(t)
\!\! &=& \!\!
\left \langle e^{i S(t)} e^{-i S} \right \rangle 
\\
 &=&\exp\left[\frac{\cos^2 \chi}{\pi}\int_{0}^{\infty} d\omega
              \frac{N(\omega)}{\omega^2}\right.\nonumber\\
       &\times & \left.\left([1-\cos(\omega t)]
              \coth(\beta \omega/2)-i \sin(\omega t)\right)\right]\nonumber\, ,
\end{eqnarray}
It causes a broadening in the energy dependence of 
the tunneling rate by $\sim \! \sqrt{\langle x(t)^2\rangle}$. 
However, we will always operate in the regime 
$eV-2\Delta \! \gg \! \cos \xi \, \sqrt{\langle x(t)^2\rangle}$, 
therefore the broadening effect can be neglected. 
Further, since the bias voltage is much larger than the energy difference, 
which is of the order of the charging energy $E_C$, 
$eV \gg |E_{\downarrow, n}-E_{1,n}|$, the transition rate is approximately, 
\begin{eqnarray}
\Gamma_{| \downarrow, n \rangle \leftarrow |1, n \rangle}
\approx 
\cos^2 (\xi/2) \, 
\Gamma_{\rm qp}
\, , 
\;\;\;\;
\Gamma_{\rm qp}=G_{\rm qp} \, V 
\, ,
\label{eqn:gamma_pump}
\end{eqnarray}
where
$G_{\rm qp}
=
2 \pi |t^{\rm qp}|^2 \rho_0^2
$ is the quasiparticle tunnel conductance.

The crucial part in our theory is the photon absorption and emission rates 
induced by $\tilde{V}_{\rm g}$. 
Although these rates are governed by the charge fluctuations $x$, 
it turns out that an incoherent pumping
 by quasiparticle tunneling also creates additional dephasing and
 modifies these rates considerably. 
Therefore, we consider second order perturbation theory in $g \sin \xi$, 
but infinite order in powers of the quasiparticle tunneling rate
$\Gamma_{\rm qp}$. 
The photon absorption rate is given by, 
\begin{eqnarray}
\Gamma_{| \uparrow, n \rangle \leftarrow |\downarrow, n+1 \rangle}
\!\! &=& \!\!
2 g^ 2 \sin^2 \! \xi 
\, {\rm Re} \!\! 
\int_0^\infty \!\!\!\! dt \, 
C_{\rm g}(t) \, 
f^{n+1}_n \, 
\nonumber \\
& & \times
i 
\pi^{|\uparrow,n \rangle}_{|\downarrow,n+1 \rangle}(t)
\, .
\label{eqn:gamma_absorp}
\end{eqnarray}
The matrix element reads 
$
f^{n+1}_n
\!=\!
|\langle n \!+\! 1| {\rm e}^{i S'} a^\dagger {\rm e}^{i S'}|n \rangle|^2
\approx 
n+1$ 
up to the order of $(g/\omega_0)^2$. 
From Eq.~(\ref{eqn:gamma_absorp}),  we can see that
 the correlator $C_{\rm g}(t)$ causes a broadening of the resonance of SSET and oscillator. 
Therefore we will get direct information about the photon emission from the correlator.

\begin{figure}[t]
\includegraphics[width= .7 \linewidth]{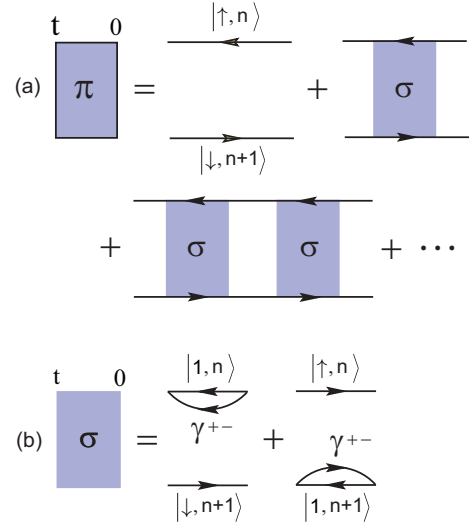}
\caption{
(a) 
The propagator describing the time evolution of the off-diagonal component of the reduced density matrix 
$\langle \uparrow,n| \rho |\downarrow,n+1 \rangle$. 
(b) 
The leading diagrams for the self-energy. 
The two diagrams are second order expansions in $\tilde{V}_{\rm qp}$. 
}
\label{fig:dephasing}
\end{figure}

The propagator $\pi^{|\uparrow,n \rangle}_{|\downarrow,n+1 \rangle}$ 
describes the time evolution of the off-diagonal component of 
the reduced density matrix $\langle \uparrow,n| \rho |\downarrow,n+1 \rangle$, 
i.e. the dephasing. 
For zeroth order, it simply oscillates with frequency $\delta \omega$ as  
$
\pi^{|\uparrow,n \rangle}_{|\downarrow,n+1 \rangle}(t) \!=\! -i {\rm e}^{i \delta \omega t}
$. 
One should keep in mind that in our system $\Gamma_{\rm qp}$
 is the strength of an incoherent pumping, that creates an additional dephasing. 
We have to consider the broadening effects by this additional 
dephasing when the correlator $C_{\rm g}(t)$ decays weakly or
 does not decay as is the case when we later relate the
 environment with another mode in an electromagnetic resonator. 
The dephasing can be treated conveniently 
using the real-time diagrammatic technique~\cite{Schoeller,Shnirman,Utsumi}. 
The propagator, diagrammatically expressed as fig.~\ref{fig:dephasing}~(a), 
is written in the Fourier space as, 
\begin{eqnarray}
\pi^{|\uparrow,n \rangle}_{|\downarrow,n+1 \rangle}(\omega)
=
\frac{1}
{\omega+\delta \omega
-
\sigma^{|\uparrow,n \rangle}_{|\downarrow,n+1 \rangle}(\omega)}
\, .
\end{eqnarray}
The self-energy can be expanded again in powers of 
$\tilde{V}_{\rm g}$, $\tilde{V}_{\rm ch}$ and also 
$\tilde{V}_{\rm qp}$. 
We account for the leading terms of the expansion in $\tilde{V}_{\rm qp}$,
 corresponding to two diagrams in fig.~\ref{fig:dephasing}~(b), 
which actually cause the dephasing~\cite{Utsumi}; 
\begin{eqnarray}
\sigma^{|\uparrow,n \rangle}_{|\downarrow,n+1 \rangle}
\!\! & \approx & \!\!
\left (
\sin^2 \xi \, 
\gamma^{+-}(t)
{\rm e}^{i (E_{\uparrow}-E_1-eV) t}
C_{\rm g}(t)^{1/4}
+\cos^2 \xi
\right.
\nonumber \\
& &
\left.
\times
\gamma^{+-}(-t) 
C_{\rm g}(-t)^{1/4}
{\rm e}^{i (E_{\downarrow}-E_1-eV) t}
\right)
\nonumber \\
& & 
\times 
{\rm e}^{i (\delta \omega+\Delta E/2) t}
\theta(t)
+
{\cal O}(g^2/\omega_0^2)
\, .
\end{eqnarray}
As we discussed when we derive Eq.~(\ref{eqn:gamma_pump}), 
the charge noise does not influence quasiparticle tunneling 
for large bias voltage $eV$. 
Then the self-energy can be approximated as 
${\rm Im}\, \sigma^{|\uparrow,n \rangle}_{|\downarrow,n+1 \rangle}(-\delta \omega)
\approx 
-\Gamma_{\rm qp}/2$, 
which is independent of the gate charge $\delta N_{\rm G}$. 
By neglecting the real part, which gives the renormalization
of the frequency $\delta \omega$, we obtain the approximate form, 
\begin{eqnarray}
\pi^{|\uparrow,n \rangle}_{|\downarrow,n+1 \rangle}(t)
\approx 
- i 
{\rm e}^{i \delta \omega t -\Gamma_{\rm qp} t /2}. 
\end{eqnarray}
within this approximation, we obtain, 
\begin{eqnarray}\label{eq_uptodown_and_photon_creation}
\Gamma_{| \uparrow, n \rangle \leftarrow |\downarrow, n+1 \rangle}
\!\! & \approx & \!\!
(n+1) \, S_{\rm g}(\delta \omega) 
=
2 (n+1) \, g^ 2 \sin^2 \! \xi 
\nonumber \\
& & \times 
\int_0^\infty \!\!\!\! dt \, 
C_{\rm g}(t) \, 
{\rm e}^{i \delta \omega t -\Gamma_{\rm qp} \, t /2} 
\, , 
\end{eqnarray}
where the dephasing effects of the incoherent pumping
can be simply included via an exponential decay with the decay rate $\Gamma_{\rm qp}/2$.

Within the same approximation, the rate of relaxation from 
$| \uparrow,n \rangle$ state to $| \downarrow,n \rangle$ state 
caused by $\tilde{V}_{\rm ch}$ is calculated as, 
\begin{eqnarray}
\Gamma_{| \downarrow, n \rangle \leftarrow |\uparrow, n \rangle}
\!\! &=& \!\!
2 \sin^2 \! \xi 
\, {\rm Re} \!\! 
\int_0^\infty \!\!\!\! dt \, 
C_{\rm ch}(t) \, 
C^n_{n,1} \, 
i \pi^{|\uparrow,n \rangle}_{|\downarrow,n \rangle}(t)
\, ,
\nonumber \\
\end{eqnarray}
where the correlation function reads (Appendix.~A), 
\begin{eqnarray}
\label{eq_chcorel}
C_{\rm ch}(t) 
\!\! &=& \!\!
\left \langle 
\left( e^{-i S/2} x e^{-i S/2} \right)(t)
\left( e^{i S/2} x e^{i S/2} \right)(0)
\right \rangle
\nonumber \\
\!\! &=& \!\!
C_{\rm g}(t)\sin^2\xi
\left[
\langle x(t)x(0) \rangle 
\!-\! 
\cos^2 \! \xi \langle p(t)x(0) \rangle^2 
\right]
,~~
\nonumber \\
\end{eqnarray}
and
\begin{eqnarray} 
\label{eq_cheff}
\langle x(t) p(0) \rangle 
=
i
\int \!\! \frac{d\omega}{2 \pi}
\frac{
N(\omega) n^-(\omega) 
}{\omega}
{\rm e}^{-i \omega t}
\, . 
\end{eqnarray}
Up to the order $(g/\omega_0)^2$ the relaxation rate becomes,
\begin{eqnarray}\label{eq_Gamma_uptpdown_via_charge_noise}
\Gamma_{| \downarrow, n \rangle \leftarrow |\uparrow, n \rangle}
\!\! & \approx & \!\!
S_{\rm ch}(\Delta E)
=
2  
\nonumber \\
& & \times
{\rm Re} \!\! 
\int_0^\infty \!\!\!\! dt \, 
C_{\rm ch}(t) \, 
{\rm e}^{i \Delta E t -\Gamma_{\rm qp} \, t /2} 
\, ,
\end{eqnarray}
The contribution of this term is often small because of rapid 
oscillations with frequency $\Delta E$, 
which is of the order of $\omega_0$
and larger than any other energy scale but $eV$. 
Therefore the integral is convergent for $t \gg 1/\Delta E$.

 The effects of the environment are contained in two noise correlators, 
 $C_{\rm ch}$ and $C_{\rm g}$ [eq. (\ref{eq_gcorel}) and (\ref{eq_chcorel})]. 
 Both correlators depend on the rotation of the qubit. 
 At the degeneracy point, $\xi=\pi/2$, they are reduced to, 
 \begin{eqnarray}
 C_{\rm g} &=& 1, \\
 C_{\rm ch} &=& \langle x(t) x(0) \rangle .\nonumber
 \end{eqnarray}
 The first result means that the charge qubit is protected against dephasing. 
 The second result means that the relaxation from 
 $|\uparrow \rangle$ state to 
 $|\downarrow \rangle$ state is caused by the charge fluctuations. 
 We will discuss these correlators 
 and the corresponding spectral functions, $S_{\rm g}(\delta\omega)$ and $S_{\rm ch}$
 defined in eq. (\ref{eq_uptodown_and_photon_creation})
 and (\ref{eq_Gamma_uptpdown_via_charge_noise}) respectively,
 for explicit choices of noise spectra later.

 The spectral functions tell us at which energy absorption
 (emission) is most effective, however one should note that
 the effect each of these spectra has on the
 lasing properties of the system is markedly different. 
 The spectral function $S_{\rm g}(\delta \omega)$ describes
 the dephasing effects, which lead to an overall
 broadening of the resonance between oscillator and SSET. 
 The corresponding operator $\sigma_- a^{\dag}$
 flips the qubit state, and creates a photon at the
 same time. Therefore
 the effectiveness of the creation of
 photons is proportional to the size of $S_{\rm g}(\delta \omega)$. 
 Only for large $S_{\rm g}(\delta \omega)$ we will have a significant number of photons in the oscillator.
 In contrast to this $S_{\rm ch}(\Delta E)$
 has a detrimental effect on the creation of photons. 
 The relaxation operator $\sigma_-$ allows transitions between the states $|\uparrow\rangle$ and $|\downarrow\rangle$. 
 If therefore $S_{\rm ch}(\Delta E)$ is large we have another efficient channel,
 that closes the lasing cycle shown in fig. \ref{fig:cycle} and only few photons will be created.

\subsection{Master equation}
\label{sec_MasterEq}

To find the density matrix in the stationary limit, we describe the system by a master equation. 
We expand the time evolution in orders of the coupling to the reservoirs. 
Dissipation and Quasiparticle tunneling can be considered to be in the Born-Markov limit,
 therefore we truncate the expansion at lowest non-zero order. 
For the charge noise we can truncate at lowest order and perform the Markov approximation for
\begin{eqnarray}
S_{\rm g}(\delta \omega)<\Gamma_{\rm qp}\,,\,\, S_{\rm ch}(\Delta E)<\Delta E \, .  
\end{eqnarray}
Then the Master equation reads, 
\begin{eqnarray}\label{eq:masterrhoincoherent}
\dot{\rho}
=
-i \, [H_{\rm sys},\rho]
+
(L_{\rm g}+L_{\rm ch}+L_{\rm qp}+L_{\rm diss}) 
\, \rho 
\, , 
\end{eqnarray}
where the Lindblad operator is separated into 4 parts. 
The coupling between oscillator and SSET is described by
\begin{eqnarray}\label{eq_Lindblad_Photon_creation}
L_{\rm g}  \rho
&=&
\frac{S_{\rm g}(\delta\omega)}{2}
\left(
2a\sigma_+\rho\sigma_- a^{\dag}- \sigma_- a^{\dag} a\sigma_+\rho
-\rho \sigma_- a^{\dag}a\sigma_+
\right)
\nonumber\\ 
& &
+\frac{S_{\rm g}(-\delta\omega)}{2}
\left(
2\sigma_- a^{\dag}\rho a\sigma_+ - 
a\sigma_+ \sigma_- a^{\dag}\rho 
\right.
\nonumber \\ 
& & 
\left.
- \rho a\sigma_+ \sigma_- a^{\dag} \right)
\,. 
\end{eqnarray}
Qubit relaxation is contained in the super-operator $L_{\rm ch}$, which reads
\begin{eqnarray}\label{eq_Lindblad_qubit_decay}
L_{\rm ch} \, \rho
\!\! &=& \!\!
\frac{S_{\rm ch}(\Delta E)}{2}
\left(2\sigma_+\rho\sigma_- - \sigma_- \sigma_+\rho
-\rho \sigma_- \sigma_+\right)
\nonumber\\ 
\!\! &+& \!\!
\frac{S_{\rm ch}(-\Delta E)}{2}
\left(2\sigma_- \rho \sigma_+ - 
\sigma_+ \sigma_- \rho-\rho \sigma_+ \sigma_- \right)
.~~
\end{eqnarray}
Incoherent pumping processes by the quasiparticle tunneling is contained in, 
\begin{eqnarray}\label{eq_Lindblad_pumping_cycle}
L_{\rm qp}\rho
&=&
\frac{\Gamma_{\rm qp}}{2} \sum_{n=0,1} (2P_{n+1,n}^{\dag}\rho P_{n+1,n} 
\nonumber \\ & &   
-P_{n+1,n}P_{n+1,n}^{\dag}\rho-\rho P_{n+1,n}P_{n+1,n}^{\dag})
\, ,~~
\end{eqnarray}
and dissipation in the oscillator part reads, 
\begin{eqnarray}
L_{\rm diss}\rho
&=&
\frac{\kappa}{2} 
\left(2a\rho a^{\dag}-a^{\dag}a\rho-\rho a^{\dag}a\right)
\end{eqnarray}
Here the Hamiltonian $H_{\rm sys}$ and the Lindblad operator do not mix the 
off-diagonal and diagonal components of the reduced density matrix. 
Therefore, the system is described by the diagonal components and evolves stochastically. 
In the basis $|\sigma,n\rangle$ we can than write the equations of motion as 
\begin{eqnarray}\label{eq_Full_Eq_of_Motion}
   \dot{\rho}_{1,n}& =&\Gamma_{\rm qp} \sin^2(\xi/2)\rho_{\uparrow,n}
                        +\Gamma_{\rm qp}\cos^2(\xi/2)\rho_{\downarrow,n}\, ,
                      \nonumber \\
   & &+\kappa(n+1)\rho_{1,n+1}-(\kappa n+\Gamma_{\rm qp})\rho_{1,n} \\
   \dot{\rho}_{\uparrow,n}& =& \Gamma_{\rm qp} \cos^2 (\xi/2)\rho_{1,n} 
   +S_{\rm g}(\delta \omega)(n+1)\rho_{\downarrow,n+1}\nonumber\\
   & &+S(-\Delta E)\rho_{\downarrow,n}+\kappa(n+1)\rho_{\uparrow,n+1}
        -\left(\Gamma_{\rm qp}\sin^2  (\xi/2) \right.\nonumber\\
   & &       
\left. + S_{\rm g}(\delta \omega)(n+1)+S(\Delta E)+\kappa n\right)\rho_{\uparrow,n}\, , \nonumber\\
   \dot{\rho}_{\downarrow,n} &=& \Gamma_{\rm qp} \sin^2(\xi/2)\rho_{1,n}
                                 +S_{\rm g}(\delta \omega) n\rho_{\uparrow,n-1}\nonumber\\
   & & +S_{\rm ch}(\Delta E)\rho_{\uparrow,n}+\kappa(n+1)\rho_{\downarrow,n+1}
       -\left(\Gamma_{\rm qp}\cos^2(\xi/2)\right.\nonumber\\
   & &    \left.+S_{\rm g}(\delta \omega)(n+1)+S(-\Delta E)+\kappa n\right)\rho_{\downarrow,n}\, ,\nonumber
  \end{eqnarray}
where we used the notation $\rho_{\sigma,n}=\langle n, \sigma|\rho|\sigma. n\rangle$.
As we will show in section \ref{sec_one_over_f_noise}, we can rewrite this master equation using standard
methods and obtain a form which is formally equivalent to the equation of motion for a laser \cite{Scully}.

In the next three sections we will discuss the stationary solutions 
of the master equation, first for low frequency noise (see
eq.~(\ref{eq_LowFrequencyImpedance})), than for coupling to a spurious resonance (see eq.~(\ref{eq:SinglemodeImpedance}))
and at last for 1/f-noise.  One should note here
 that we assume that 1/f-noise can be described as Gaussian and ergodic noise,
 which is not necessarily true and depends on the microscopic model \cite{Schriefl2005,Gerd_Juan_recent}.
We discuss the distribution probability of photons, 
\begin{eqnarray}
\rho_n=\sum_{\sigma=\uparrow,\downarrow,1} 
\langle \sigma,n| \rho  |\sigma, n\rangle \, , 
\end{eqnarray}
the average number of photons 
$\langle n \rangle = \sum_n n \rho_n$, etc. 
We will also investigate the current through the SSET, which is given by
\begin{eqnarray}
I=
e \Gamma_{\rm qp}\left( \rho_1 
+
 \sin^2\frac{\xi}{2} \rho_{\uparrow}
+ 
\cos^2\frac{\xi}{2} \rho_{\downarrow}\right)
\, ,
\end{eqnarray}
where the density matrix for the qubit states is obtained by tracing out photon number state as, 
\begin{eqnarray}
\rho_\sigma=\sum_n \langle \sigma,n| \rho  |\sigma, n\rangle \, .
\end{eqnarray}

\section{The double-dot}

In a double dot we consider a situation where a single charge can either sit on the left or right dot or both dots are unoccupied.
Therefore we can write the double-dot Hamiltonian as,
\begin{eqnarray}
 H_{qbit}&=&\epsilon_L |1,0 \rangle\langle 1,0|+\epsilon_R|0,1\rangle\langle 0,1|+\epsilon_0 |0,0\rangle\langle 0,0|\nonumber \\ 
  & & + t\left(|1,0 \rangle\langle 0,1|+|0,1 \rangle\langle 1,0|\right)
\end{eqnarray}
Here the eigenstates of the double-dot are given by $|1,0\rangle$ (left dot occupied), $|0,1\rangle$ (right dot occupied)
and $|0,0\rangle$ (both dots unoccupied). To map this case to the SSET we replace the states of the SSET with the double dot states
using the following rules, $|0,1\rangle\rightarrow |N=2\rangle$, $|1,0\rangle\rightarrow |N=0\rangle$ and $|0,0\rangle \rightarrow |N=1\rangle$.
All energies have to be mapped accordingly. The coupling between the dots is equivalent to the Josephson coupling, $t=E_J/2$.
A detailed discussion of the lasing cycle in the case of the couple dot can be found in Refs. \onlinecite{Marthaler_Jin_DDotLasing,Marthaler_Jinshuang_DDotLasing}.

For the double dot it is also reasonable to diagonalize the subspace which contains the charge states $|1,0\rangle$ and $|0,1\rangle$. This 
gives us states which act in an equivalent fashion to the states defined in eq. (\ref{eq_Upanddownarrow_total_states_at_beginning}),
\begin{eqnarray}
|\! \uparrow \rangle   &=& \cos\frac{\xi}{2}|1,0\rangle + \sin\frac{\xi}{2}|0,1\rangle\, ,\\
|\! \downarrow \rangle &=& \sin\frac{\xi}{2}|1,0\rangle - \cos\frac{\xi}{2}|0,1\rangle\, ,\nonumber\\
\tan\xi &=& 
\frac{2t}{\epsilon_R-\epsilon_L} \, .\nonumber
\end{eqnarray}
This allows us to write the Hamiltonian of the double dot in the form,
\begin{eqnarray}\label{eq_SSET_in_the_form_of_a_twolevelsystem}
H_{\rm qbit}
&=&
\frac{
\Delta E \sigma_z
}{2} 
+
\epsilon_0 
\, |0,0\rangle\langle 0,0|
\, ,
\\
\Delta E &=& 
\sqrt{4t^2+(\epsilon_L-\epsilon_R)^2}, \nonumber
\end{eqnarray}
The terms in the master equation, which effect the photon creation (\ref{eq_Lindblad_Photon_creation})
and decay (\ref{eq_Lindblad_qubit_decay}) can then be adapted directly for the states $|\! \uparrow \rangle$ and $|\! \downarrow \rangle$ of the 
double dot. The pumping rates, which create the population inversion, are described by eq. (\ref{eq_Lindblad_pumping_cycle}). These terms can 
be adapted for the double dot, using the substitution rules discussed above.

\section{The electromagnetic environment}

In this section we will consider two examples that have been considered previously
 within the context of tunneling in small junctions \cite{SingleChargeTunneling}. The
 first case is coupling to a large ohmic impedance. We have chosen this example because it gives results
 similar to 1/f noise. In fact,
 as we will show in section \ref{sec_one_over_f_noise}, one can see 1/f noise 
 as the classical limit
 of a large ohmic impedance. The second example is coupling to a single mode in equilibrium.
 Effects of this have been measured for a single charge device with a well characterized
 environment e.g. in Ref. \onlinecite{Lindell2003}. 

\subsection{High Impedance Environment}\label{subsec_LowFrequencyNoise}

\begin{figure}
  \includegraphics{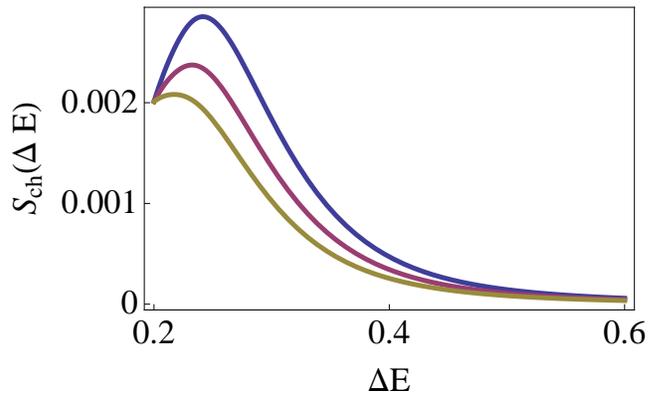}
\caption{The correlator $S_{\rm ch}(\Delta E)$ as a function of the
          energy splitting $\Delta E$ for a fixed noise width $\sqrt{k_B T \epsilon_C}=0.07$
(all energies are given in units of the charging energy $E_C$).
(blue) $\epsilon_C=0.05$, (magenta) $\epsilon_C=0.03$,
         (brown) $\epsilon_C=0.01$.
         For the parameters  : $E_J=0.2$, $\Gamma_{\rm qp}=0.0325$.
 }\label{fig_sxlowrequency}
\end{figure}

 If the qubit is coupled to an ohmic resistor the effective impedance in eq. (\ref{eq_N_of_omega_as_function_of_Z})
 takes the form \cite{SingleChargeTunneling}
 \begin{eqnarray}
  \frac{{\rm\, Re} Z(\omega)}{R_K}=\frac{\epsilon_C \omega_R}{\omega^2+\omega_R^2}
 \end{eqnarray}
  where $\omega_R=1/RC_{{\rm eff}}$ is the RC-cutoff frequency, $\epsilon_C=e^2/C_{\rm eff}$
  and $C_{\rm eff}$ is a capacitance that depends on the details of the coupling to the external resistor
  with ohmic resistance $R$. We consider noise sharply peaked for low frequencies, such that 
  $\omega_R\ll k_B T$.  In this case the dephasing correlator becomes,
 \begin{eqnarray}
  C_{\rm g}(t)=e^{\frac{\epsilon_C \cos^2\xi (k_B T-i\omega_R)(1-e^{-\omega_R t})}{2\omega_R^2}
             -\frac{\epsilon_C k\cos^2\xi k_B T t}{2\omega_R}}\, .
 \end{eqnarray}
 The decay time can not be larger than $t=1/\Gamma_{\rm qp}$. Therefore we can expand the exponent in this correlator
 in the short time limit for $\omega_R\ll \Gamma_{\rm qp}$ and get 
\begin{equation}\label{eq_Cgforlowfrequencynoise}
  C_{\rm g}(t)=\exp\left(-\epsilon_C\cos^2 \xi\left[i t+k_B T t^2\right]\right)
 \end{equation}
  While we will not explicitly consider the opposite limit $\omega_R \gg \Gamma_{\rm qp}$, one should note that in this case
  our theory reproduces the standard quantum optics results expected for a smooth spectral noise density.
  However, we will discuss this transition for small charge noise strength $\epsilon_C \ll \Gamma_{\rm qp}$.

 The limit considered here  corresponds to a large resistor, $R\rightarrow \infty$, in the 
 effective impedance \cite{SingleChargeTunneling},
 \begin{eqnarray}\label{eq_LowFrequencyImpedance}
  {\rm Re}Z(\omega)/R_K \approx \epsilon_C\delta(\omega)\, .
 \end{eqnarray}

\begin{figure}
\includegraphics[width =3 in]{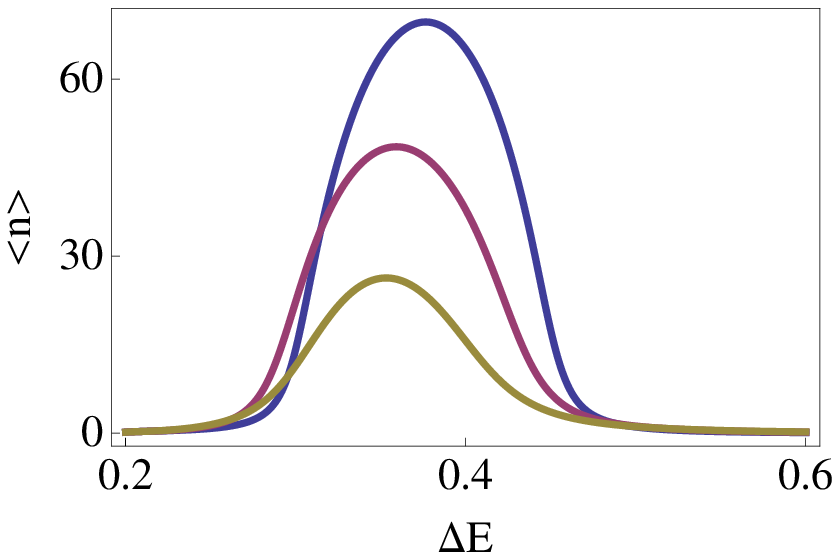}
\includegraphics[width =3 in]{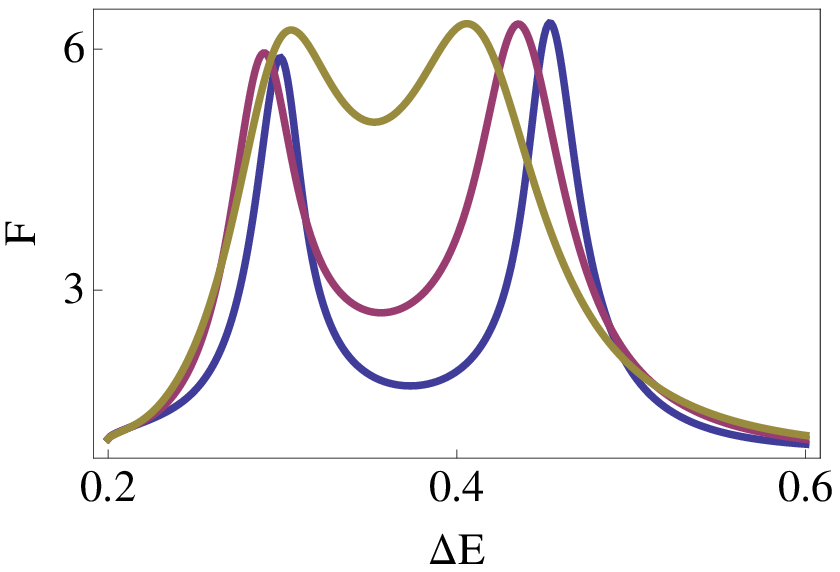}
 \caption{The average number of photons
 $\langle n \rangle$ and the Fano-Factor as a function of the 
 energy splitting $\Delta E$ (all energies are given
 in units of the charging energy $E_C$). (blue)  $\epsilon_C=0.005$,
 (magenta) $\epsilon_C=0.01$, (brown) $\epsilon_C=0.015$.
 For the parameters: $E_J=0.2$, $k_B T=0.5$, $\omega_0=0.4$,
 $\Gamma_{\rm qp}=0.0325$, $\kappa=0.0006$,
 $g=0.004$.}\label{fig_nforlowfrequency}
\end{figure}

 The dephasing spectral function at the 
 energy $\delta \omega$ becomes
 \begin{eqnarray}\label{eq_Sg_for_low_frequencies}
  S_g(\delta \omega)&=&\frac{1}{2}\sqrt{\frac{\pi}{\epsilon_C \cos^2\xi k_B T}}
  \, e^{\frac{\left(\Gamma_{\rm qp} - i (\delta\omega+\epsilon_C\cos^2\xi)\right)^2}{4 \tilde{\epsilon}_C k_B T}}\nonumber\\
& & \times         {\rm erfc}\left[\frac{\Gamma_{\rm qp} - 
       i (\delta\omega+\epsilon_C\cos^2\xi)}{2 \sqrt{\epsilon_C \cos^2\xi \,k_B T}}\right] 
     + {\rm c.c.} \nonumber\\
 \end{eqnarray}
 where ${\rm erfc}[x]$ is the complementary error function.
 We can immediately analyze two important limits of this spectral function,
 \begin{eqnarray}\label{eq_SgApproxforLowFrequency_in_two_limits}
   S_g(\delta \omega)&\propto &\left\{\begin{array}{cc}
         e^{-\frac{(\delta\omega+\epsilon_C\cos^2\xi)^2}{4\epsilon_C \cos^2\xi\,k_B T}} 
                 &,  \,\,     \cos\xi\sqrt{ \epsilon_C  k_B T}\gg \Gamma_{\rm qp} \\
        \frac{2 \Gamma_{\rm qp}}{\Gamma_{\rm qp}^2+\delta\omega^2}
                      &, \,\,     \cos\xi\sqrt{ \epsilon_C k_B T}\ll \Gamma_{\rm qp}
                        \end{array}
\right.\nonumber\\
 \end{eqnarray}
 If we have strong noise
 we have a Gaussian spectral function with a width 
 defined by $\cos\xi\sqrt{\epsilon_C k_B T}$.
 The Gaussian function  is peaked 
 at $\Delta E-\omega \approx\cos\xi\epsilon_C$. 
 This corresponds to the fact, that in each qubit flip,
 a photon is created and additionally the energy $\epsilon_C\cos\xi$ is transfered to the environment.
 Previoulsy it has been shown that in a similar situation, such a
 shift of the resonance can be used to create inversionless lasing \cite{Marthaler_Inversionless}
 or enhanced photon emission in a double dot \cite{Peta_Theory_enviornment_enhanced_emission}.
 
 Note that the width and center of $S_g(\delta \omega)$ depends on the rotation of the qubit.
 Directly at the degeneracy point, $\delta N_g=0$ , we have $\cos\xi=0$ 
 and therefore $S_g$ is simply a Lorentzian with the width being determined
 by $\Gamma_{\rm qp}$. Overall we see that for strong low frequency noise we get,
 as one would expect, a Gaussian resonance peak and for weak noise we get Lorentzian
 resonance peak. 

 The spectral function $S_{\rm ch}(\Delta E)$
 is proportional to the relaxation rate. However, for any given configuration of our system 
 this correlator will be evaluated for energies larger than
 the Josephson coupling, $\Delta E \geq E_J$. Generally for low frequency noise this means
 that the correlator has already decayed for relevant frequencies and can be ignored.
 However, one should note that we have a significant broadening, 
 because of the dephasing caused by $  \Gamma_{\rm qp} $, and as we move
 away from degeneracy, $\cos \xi>0$, we get an additional broadening through $\cos\xi\sqrt{\epsilon_C k_B T}$
 and an overall frequency shift $\epsilon_C \cos\xi$ (see eq. (\ref{eq_Cgforlowfrequencynoise})). 
 In fig. \ref{fig_sxlowrequency} we can see a plot of the correlator $S_{\rm ch}(\Delta E)$.
 For this plot we have chosen a fixed value for the noise width $\sqrt{\epsilon_C k_B T}$
 and than changed the coupling strength $\epsilon_C$. One can see that for large coupling
 the frequency shift has a significant impact and the spectral function has a peak for 
 $\Delta E>E_J$. However, if the width is mostly caused by the noise temperature
 the correlator decays monotonically.

 That means that the effect of transverse coupling for 
 low frequency noise is only important for a very strong coupling to the environment.
 Therefore in our discussion of 1/f noise 
 (see sec. \ref{sec_one_over_f_noise}), where we make the formal
 transition to the classical high temperature limit, we can ignore transverse coupling because we keep $\epsilon_C$ 
 and therefore the frequency shift small.

 In fig. \ref{fig_nforlowfrequency} we show the average number of photons for high
 ohmic impedance, calculated by numerically solving eq. (\ref{eq_Full_Eq_of_Motion})
 in the stationary limit. As predicted from eq. (\ref{eq_SgApproxforLowFrequency_in_two_limits})
 we see that the resonance peak changes from a Lorentzian to a more Gaussian form for increasing noise coupling 
 $\epsilon_C$. Nonetheless, we still have a significant 
 Lorentzian component given by $\Gamma_{\rm qp}$. 
 One should also note that the maximum shifts to energies smaller than the oscillator
 frequency, $\Delta E<\omega$. The reason for this is the decrease of the coupling strength as we tune
 away from the degeneracy point. We will discuss this in more detail in sec. \ref{sec_one_over_f_noise}.  
 We also show the Fano-Factor which 
 shows the standard behavior for a laser. When we start to populate the oscillator,
 the distribution has a thermal shape, and therefore the Fano-Factor is large. As we move closer to resonance the 
 number of photons grows and the distribution takes a form similar to a Poisson distribution, which means the Fano-Factor 
 moves closer to one.

\subsection{Coupling to a single mode}
\label{subsec_Couplingtoasinglemode}

 In the limit of coupling to a single mode, the effective impedance is reduced to \cite{SingleChargeTunneling},
 \begin{eqnarray}\label{eq:SinglemodeImpedance}
  {\rm Re} Z(\omega)/R_K=\epsilon_C [\delta(\omega-\omega_L)+\delta(\omega+\omega_L)]\, .
 \end{eqnarray}
 In this case the dephasing spectral function becomes \cite{SingleChargeTunneling}
 \begin{eqnarray} \label{eq_Sgforsinglemode}
  S_{\rm g}(\delta\omega) &=& \exp\left(-[\eta_+ +\eta_-]\right)\\
   & &\times\sum_{n,m}\frac{\eta_+^n\eta_-^m}{n! m!}
  \frac{2\Gamma_{\rm qp}}{\Gamma_{\rm qp}^2+(\delta\omega+(n+m)\omega_L)^2}\nonumber\, ,\\
 \eta_{\pm} &=& \frac{\epsilon_C \cos^2\xi}{\omega_L}\frac{\pm 1}{e^{\pm\beta \omega_L}-1}\nonumber\, .
 \end{eqnarray}
 This gives us several peaks in the photon number. However one should note that the largest peak is still 
 located at $\Delta E-\omega_0=0$. Additionally we have peaks at $\Delta E-\omega_0= \bar{m}\omega_L$, with 
 $\bar{m}=\pm 1,\pm 2 \ldots$. For small temperatures, $\beta\rightarrow \infty$, 
 the environment can only absorb energy and we only have peaks at $\bar{m}\geq 0$.

  The relaxation spectral function can be be written in a compact way by defining
 \begin{eqnarray}
  S_{\rm ch,eff}(\omega)&=&\frac{2 \Gamma_{\rm qp} \omega_L^2}{\pi^2}\tan^2 \xi\left(
 \frac{\pi\eta_-}{\Gamma_{\rm qp}^2+(\omega-\omega_L)^2}\right.\\
   &+&\frac{\pi\eta_+}{\Gamma_{\rm qp}^2+(\omega+\omega_L)^2}
    +\frac{\eta_+^2}{\Gamma_{\rm qp}^2+(\omega+2\omega_L)^2}\nonumber\\
   &+&\left.\frac{\eta_-^2}{\Gamma_{\rm qp}^2+(\omega-2\omega_L)^2}
    -\frac{\eta_-\eta_+}{\Gamma_{\rm qp}^2+\omega^2}\right)\, .\nonumber
 \end{eqnarray}
  If we convolve this spectral function with eq. (\ref{eq_Sgforsinglemode}),
 we obtain the relaxation spectral function 
 \begin{eqnarray}\label{eq_Schforsinglemode}
  S_{\rm ch}(\Delta E)&=& \exp\left(-[\eta_+ +\eta_-]\right)\\
   & &\times\sum_{n,m}\frac{\eta_+^n\eta_-^m}{n! m!}
   S_{\rm ch,eff}(\Delta E+(n+m)\omega_L)\nonumber\, .
 \end{eqnarray}
 \begin{figure}
  \includegraphics{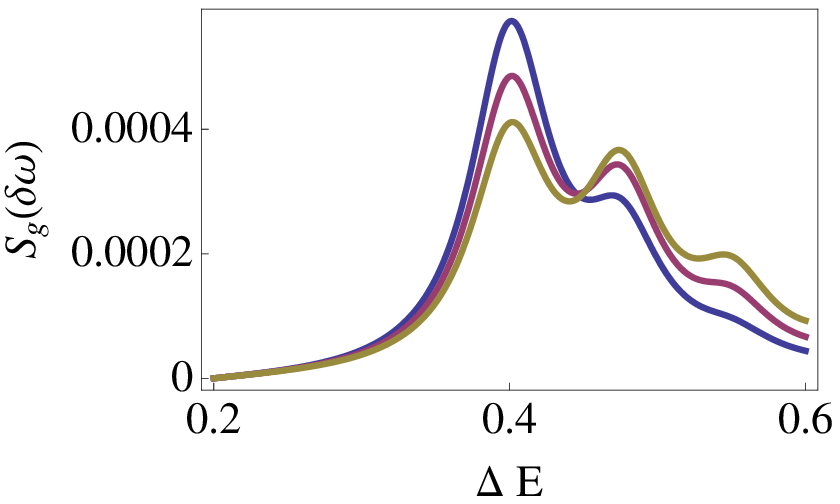}
   \includegraphics{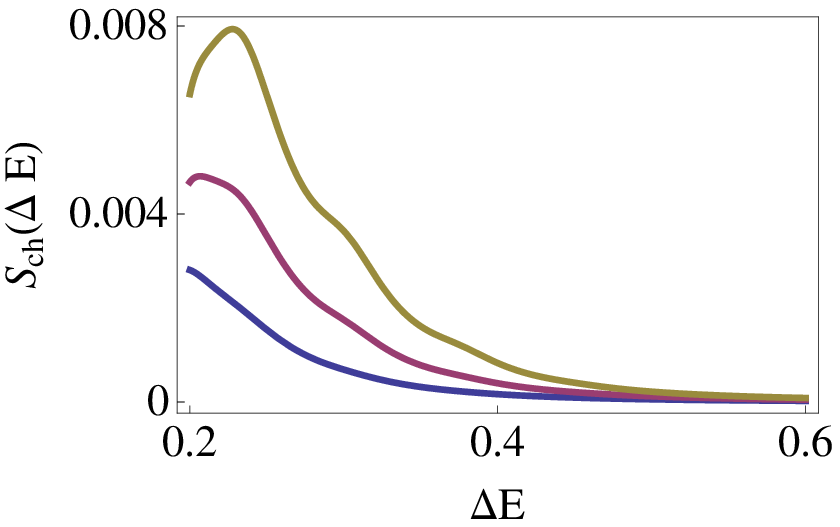}
  \caption{The correlators $S_{\rm g}(\delta \omega)$ and $S_{\rm ch}(\Delta E)$ as a function of the
          energy splitting $\Delta E$, for coupling to a single mode.
(all energies are given in units of the charging energy $E_C$).
(brown) $\epsilon_C=0.05$, (magenta) $\epsilon_C=0.075$,
         (blue) $\epsilon_C=0.1$.
         For the parameters  : $E_J=0.2$, $k_B T=0.02$, $\Gamma_{\rm qp}=0.0325$, $\omega_0=0.4$,
        $\omega_L=0.075$, $g=0.004$.
 }\label{fig_ssinglemode}
 \end{figure}
  Here we can see that the total spectral density $S_{\rm ch}$,
  will consist of multiple peaks which are repeating with distance $\omega_L$.  
  This means relaxation will be strong always for $\Delta E=\bar{m}\omega_L$, 
 with $\bar{m}=\pm1,\pm2,\ldots$. Each maximum in the relaxation rate will decrease the number of photons at that
 particular qubit splitting. At the same time incoherent Cooper-pair tunneling
 is possible and we get an additional maximum in the current. Together with the maxima
  which are caused by the lasing
 transition it is possible that we see several peaks in the current. 

 In fig. \ref{fig_ssinglemode}
 one can see an example for $S_{\rm g}(\delta \omega)$ and $S_{\rm ch}(\Delta E)$
 for the case of coupling to a single mode.
 If the system is strongly coupled ($\epsilon_C$ is large) 
  we get several resonance peaks in
 the coupling between the SSET and oscillator.
 However, these peaks decay very quickly as $\epsilon_C$
 becomes smaller. To see clearly visible peaks in
 the relaxation spectral function $S_{\rm ch}(\Delta E)$ we would have to choose
 very large $\epsilon_C$. This is because we choose a off-resonant mode $\omega_L<E_J$.

  \begin{figure}[t]
   \includegraphics[width= 7 cm]{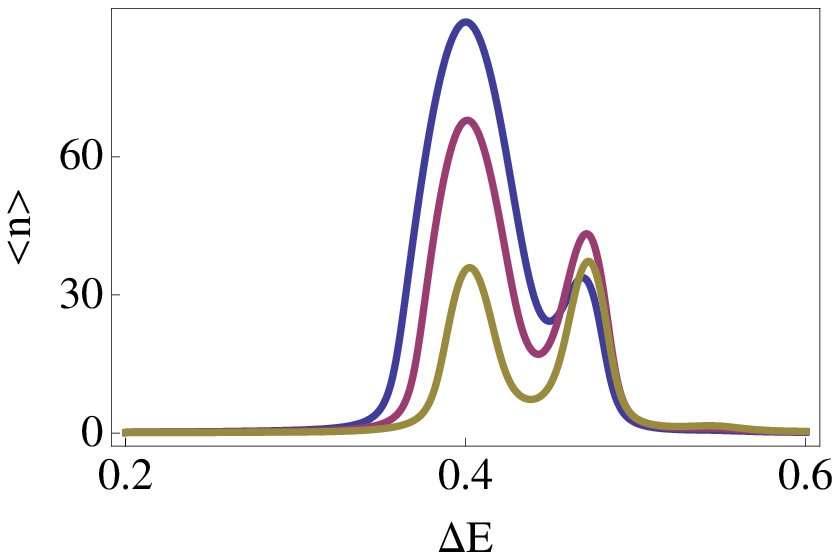}
  \includegraphics[width= 7 cm]{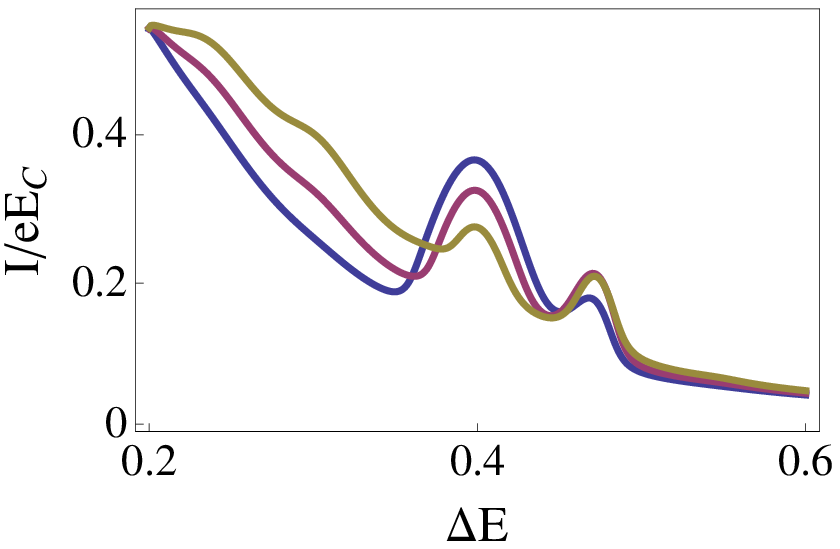}
 \caption{The average number of photons $\langle n \rangle$
 and the current $I$ as a function of the energy splitting (all energies in units of the
 charging energy $E_C$). (blue) $\epsilon_C=0.05$, (magenta) $\epsilon_C=0.075$, (brown) $\epsilon_C=0.1$.
 For the parameters: $\omega_L=0.1$, $E_J=0.2$, $g=0.004$,
 $\Gamma_{\rm qp}=0.0325$, $\kappa=0.00005$, $k_B T=0.02$, $\omega_0=0.4$.}\label{fig_lcnoise}
 \end{figure}

 Similar to the case for low frequency noise, relaxation 
 decays quickly for coupling to a single mode. But it can still have a significant impact.
 In fig. \ref{fig_lcnoise} we see the average number of photons $\langle n \rangle$ and 
 the current $I$ through the SSET as a function of the energy splitting. 
 As the coupling to the noise $\epsilon_C$ increases
 we see the maximum number of photons decrease. However, the second resonance peak for $\Delta E-\delta \omega=\omega_L$
 becomes comparable to the first resonance. For large  $\epsilon_C$ the second peak is even higher than the first peak. This is the case
 because the contribution of $S_{ch}(\Delta E)$ has decayed.
 The current shows a peak for every peak in the photon number. One can also see additional structure in the 
 current which are a result of the incoherent cooper-pair 
 tunneling caused by $S_{ch}(\Delta E)$.

\section{1/f Noise}\label{sec_one_over_f_noise}

 The existence of noise with a $1/f$ type spectrum seams to be universal to all physical systems
 and has been an object of intense study \cite{Dutta1981,Mueck2005,Schriefl2005,Schriefl2006,Gerd_Juan_recent}. 
 Generally $1/f$ noise is considered classical noise, therefore we will consider the dephasing correlator
 in the classical limit
 \begin{eqnarray}
  C_{\rm g}=\exp\left(\frac{\cos^2\xi}{\pi}\!\! \int_0^{\infty}\!\!\!d\omega\, S(\omega)\frac{\cos(\omega t)-1}{\omega^2} \right)
 \end{eqnarray}
 where we defined $S(\omega)=N(\omega)\coth(\omega/2 k_B T)$. This is equivalent to replacing the coupling 
 to a bosonic reservoir in eq. (\ref{eq_V_coupl_to _charge_noise}) with a coupling to a 
 classical fluctuating variable.  Formally, we can make the transition to the classical case
 by choosing the spectral function to be constant, $N(\omega)={\cal N}$, and the temperature to be
 large. In this case the spectral function becomes,
 \begin{eqnarray}
  S(\omega)=\frac{k_B T \cal{N}}{\omega}\, .
 \end{eqnarray}
 The factor ${\cal N} k_B T$ characterizes, 
 together with the low frequency cutoff $\omega_{\rm min}$, the strength of 1/f noise. 
 The resulting dephasing correlator reads,
\begin{eqnarray}
 C_{\rm g}=\exp\left(\frac{k_B T {\cal N} {\rm Re}\ln(\omega_{\rm min}t)}{2\pi}\,t^2\right)
 \end{eqnarray}
 The decay rate of the correlator is much larger then the cutoff frequency,
 this means that the logarithmic time dependence $\ln(\omega_{\rm min}t)$ has no impact 
 on the decay of $C_{\rm g}$. We define a new constant 
  $\epsilon_C=\frac{{\cal N}{\rm Re}\ln(\omega_{\rm min}/ \Gamma_{\rm qp})}{2\pi}$ 
  to describe the noise strength. Here we replaced the time $t$ in the $\ln$-function with the timescale given by $1/\Gamma_{\rm qp}$.
  From this we get the dephasing spectral function 
 \begin{eqnarray}
  S_g(\delta \omega)&=&\frac{1}{2}\sqrt{\frac{\pi}{\epsilon_C \cos^2 \xi k_B T}}
   e^{\frac{(\Gamma_{\rm qp} - i \delta\omega)^2}{4 \epsilon_C \cos^2 \xi
    k_B T}}\\
         & & \times {\rm erfc}\left[\frac{\Gamma_{\rm qp} - 
       i \delta\omega}{2 \sqrt{\epsilon_C \cos^2 \xi k_B T}}\right]
     + c.c. \, , \nonumber
 \end{eqnarray}
  which is the classical limit of eq. (\ref{eq_Sg_for_low_frequencies}).

 For classical low frequency noise relaxation will be rather small,
 and decay quickly, as $\Delta E$ increases (see sec. \ref{subsec_LowFrequencyNoise}).
 Additionally our dephasing spectral function is symmetric, $S_{\rm g}(\delta \omega)=S_{\rm g}(-\delta\omega)$.
 In this case we can derive an approximate analytical solution
 for the density matrix in the stationary limit. 

 We want to derive an effective equation for the 
 probability distribution of the number of photons in the
 oscillator $\rho_n=\sum_{\sigma}\rho_{\sigma,n}$. We do this by
 tracing out the degrees of freedom
 of the SSET in the equation of motion given by eq. (\ref{eq_Full_Eq_of_Motion}).
  Using the relation $\rho_n=\rho_{1,n}+\rho_{\uparrow,n}+\rho_{\downarrow,n}$ 
   and the assumption that the time scales of the SSET are faster than the time scales of the
   oscillator
   we can form a closed set of equations\, ,
   \begin{eqnarray}\label{eq:rhoclosed}
    \frac{d}{dt}\left(\begin{array}{c}
     \rho_{\uparrow,n-1}\\
     \rho_{\downarrow,n}
    \end{array}
     \right)& =& \left(
     \begin{array}{cc}
       \alpha_{1,n}& S_{\rm g}(\delta \omega) n\\
      S_{\rm g}(\delta \omega) n  & \alpha_{2,n}\\
     \end{array}\right)
     \left(\begin{array}{c}
     \rho_{\uparrow,n-1}\\
     \rho_{\downarrow,n}
    \end{array}\right)\nonumber \\
    & & +\left(\begin{array}{c}
      \beta_1 \rho_{n-1}\\
      \beta_2 \rho_{n}
      \end{array}\right)\, ,\\
      \alpha_{1,n}& =& -\frac{\Gamma_{\rm qp}(\cos 2\xi+7)}{4(\cos \xi+3)}-S_{\rm g}(\delta \omega) n \, ,\nonumber\\
      \alpha_{2,n}& =& \frac{\Gamma_{\rm qp}(\cos 2\xi+7)}{4(\cos \xi-3)}-S_{\rm g}(\delta \omega) n \, ,\nonumber\\
      \beta_1& =&  \frac{2\Gamma_{\rm qp}\cos^4 (\xi/2)}{3+\cos \xi} \, , \nonumber   \\
      \beta_2 & =& \frac{2\Gamma_{\rm qp} \sin^4 (\xi/2) }{3-\cos \xi}\, .\nonumber  
      \end{eqnarray}
    This set of equations
   can be solved in the stationary limit and we get an equation for the effect of the artificial atom on the
   oscillator
   \begin{eqnarray}\label{eq:effectivn}
    \dot{\rho}_n &=& \Gamma_{n}^{+}\rho_{n-1}
             -\left(\Gamma_{n+1}^{+}+\Gamma_n^{-}+\kappa n\right)\rho_{n}\\
          &     & +\left(\Gamma_{n+1}^{-}+\kappa (n+1)\right)\rho_{n+1}\, ,\nonumber
   \end{eqnarray}
 where $\Gamma_{n}^{+}=\Gamma_{T,n}\cos^4 (\xi/2)$ is the rate increasing
 the number of photons, $\Gamma_n^{-}=\Gamma_{T,n}\sin^4 (\xi/2)$ is the rate decreasing
 the number of photons and
 \begin{eqnarray}
  \Gamma_{T,n}=8 \Gamma_{\rm qp}
               S_{\rm g} (\delta \omega) n/\left[\Gamma_{\rm qp}(\cos \xi+7)+24 S_{\rm g} (\delta \omega) n\right].\nonumber \\
 \end{eqnarray}
  For $\cos (\xi/2)>\sin (\xi/2) $ we have a net increase of the number of photons and
  the rate $\Gamma_{T,n}$ is directly proportional to $S_{\rm g}(\delta \omega)$. 
  One should note here
  that eq. (\ref{eq:effectivn}) is formally equivalent to the equation of motion 
  as it can be derived for the same system in the case of coherent coupling using standard methods \cite{Scully}.
  In fact if we choose the limit of small charge noise, $\epsilon_C\rightarrow 0 $, 
  eq. (\ref{eq:effectivn}) is actually exactly equivalent to the standard lasing result of Ref. \onlinecite{Scully}.

  Eq. (\ref{eq:effectivn}) can be solved 
  and
  we get for the density matrix
   \begin{eqnarray}\label{eq_solrhonstrongcoupling}
    \rho_n& =& \rho_0\Pi_{i=0}^n\,\frac{A}{B+i}\, , \\
    A &=&\frac{\Gamma_{\rm qp} \cos^4 (\xi/2)}{3 \kappa}\, ,\nonumber \\
    B &=& \frac{\Gamma_{\rm qp}\sin^4 (\xi/2)}{3\kappa}+\frac{\Gamma_{\rm qp}(\cos 2\xi+7)}{24 S_{\rm g}(\delta \omega)}\, ,\nonumber 
   \end{eqnarray}
  where $\rho_0$ is a normalization constant. 

 If $\cos(\xi/2)>\sin(\xi/2)$ and $S_{\rm g}$ is not to small this 
 function has a sharp peak. In this case the average number of photons is given in good approximation
 by the position of this peak,
 \begin{eqnarray}\label{eq_numberofphotons_analytical}
 \langle n \rangle  & \approx & A-B \\
       &=& \frac{\Gamma_{\rm qp}}{3\kappa}\cos\xi-\frac{\Gamma_{\rm qp}(\cos 2\xi+7)}{24 S_{\rm g}(\delta \omega)}\, .
         \nonumber
\end{eqnarray}
 We see here that the number of photons increases
with the size of $S_{\rm g}(\delta \omega)$, since the second term in eq. (\ref{eq_numberofphotons_analytical})
decreases.  There are two key observations we can make from this analytical equation for $\langle n \rangle$.

 First  one should note that in the limit of strong 1/f-noise, $\cos\xi\sqrt{\epsilon_C k_B T}> \Gamma_{\rm qp}$,
 the resonance peak will be shifted to smaller frequencies. This is the case because of the decay of the transverse
 coupling between SSET and oscillator with the increase of $\Delta E$.
 To find the approximate position of the actual peak
 we solve $\partial_{\Delta E} S_{\rm g}(\delta \omega)=0$ for strong $1/f$ noise and get the peak position,
 \begin{eqnarray}
 \Delta E_{\rm max}\approx \frac{1}{2}\left(\omega_0 + \sqrt{\omega_0^2-16\epsilon_C k_B T}\right)\, .
 \end{eqnarray}
 This shift can also be observed for coupling to a high impedance environment [Fig. \ref{fig_nforlowfrequency}].

Now let us consider the number of photons for the qubit close to the symmetry point, $\xi=\pi/2$,
where we assume that it is in resonance with the oscillator, $\Delta E=\omega$. In this case we can expand
the number of photons around the symmetry point and get,
\begin{eqnarray}\label{eq_numberofphotons_close_to_symmetry}
 \langle n \rangle &\approx& 
        \frac{\Gamma_{\rm qp}}{3\kappa}(\xi-\pi/2)\\
       & &    -\frac{\Gamma_{\rm qp}}{4 g^2}\left(\frac{\Gamma_{\rm qp}}{2}\right.\,
              +\left.\frac{ k_B T\epsilon_C}{\Gamma_{\rm qp}}(\xi-\pi/2)^2\right)\, .\nonumber
\end{eqnarray}
 We can see here that the second, negative term in this equation decreases quadratically,
 as we move closer to symmetry. The positive first term decreases only linearly. To create coherent coupling
 between oscillator and SSET it is very important to maximize the number of photons. From
 eq. (\ref{eq_numberofphotons_close_to_symmetry}) we can immediately see that the number of photons
 has a maximum for
 \begin{eqnarray}
 (\xi-\pi/2)=\frac{g^2 \Gamma_{\rm qp}}{\kappa k_B T \epsilon_C}\, .
 \end{eqnarray}
  We see here that especially for large charge noise
  $\epsilon_C$ it can be of advantage to stay closer to the
  symmetry point. The loss in population inversion
  is compensated by the suppression of charge noise.

 \section{Conclusion}

 In this paper we discuss the single artificial atom maser which can be realized using a SSET coupled to an
 oscillator or a double dot coupled to a transmission-line resonator.
 For such a system  it is necessary
 to bias the system away from the symmetry point,
 to create population inversion, which makes it susceptible to charge noise.
 In this case it is a reasonable
 approach to describe the system using a polaron transformation.
 We have shown the transformation of the qubit Hamiltonian and the resulting new noise operators.
 The relevant noise correlators have been shown in explicit form and
 we discuss them for two relevant examples, low frequency noise and coupling to a single mode. Then we perform
 the transition to the classical case, which describes 1/f-noise.

 We find that low frequency noise creates a Gaussian resonance peak, which is shifted
 to smaller qubit energy splitting. Coupling to an additional mode can create 
 additional resonances in the fundamental mode and in the current. One should note here
 that coupling to an additional mode can even create peaks in the current if there is no excitation in the 
 oscillator.

 For our discussion of 1/f-noise we can show that our master equation is formally equivalent to the standard description of a laser \cite{Scully}.
 We can find an analytical solution for the density matrix and the average number of photons. We can then show
 that it can be of advantage to be closer to the symmetry point, especially if charge noise is strong. This has
 direct implications for experiments on mechanical oscillators coupled to an SSET. To create coherent lasing, 
 one should tune the system rather close to the symmetry point, since the reduction in population inversion
 affects the photon number only linearly, but the reduction in charge noise is quadratic.

  The model presented in this paper can also describe many other system that are currently studied. 
  An example are suspended carbon nanotubes. Electrons can hop onto a free state of the nanotube,
  and it has already been shown that coupling to the mechanical vibration of the 
  nanotube can be used to create cooling \cite{Belzig_Nanotube,Burkard_Nanotube}. Similarly lasing could be achieved. 
  Since charge is relevant degree of freedom for these systems, strong coupling to noise should be expected.

\section*{Acknowledgments}

We thank Y. Nakamura, G. Sch\"on and A. Shnirman for helpful discussions.
This work has been supported by Strategic International Cooperative Program
of the Japan Science and Technology Agency (JST) and
by the German Science Foundation (DFG).

 \section*{Appendix}
 To find an explicit form for the relaxation correlator (\ref{eq_chcorel}) we use the
 generating function,
 \begin{eqnarray}
  F=\langle T e^{i \int dt' \left(\lambda (t')  p(t')\cos\xi+\mu(t')x(t')\sin\xi\right)}\rangle\, ,
 \end{eqnarray}
 where $T$ is the time sorting operator.
  Using function we can show that the relaxation correlator is given by
  \begin{eqnarray}\label{eq_apA_Cchfromgenerator}
  C_{\rm ch}(t)=-\left. \frac{d^2 F}{d\mu(0)d\mu(t)}
    \right\rvert_{{\tiny
    \begin{array}{rl}
     \lambda(t')&=(\delta(t-t'+\eta)+\delta(t-t'-\eta))/2\\
                &-(\delta(t'+\eta)-\delta(t'+\eta))/2\\
     \mu(t')&=0
    \end{array}}}\, .\nonumber\\
  \end{eqnarray}
  The explicit form of the generating function reads,
 \begin{eqnarray}
   F &=& \exp\left[-\int dt_1 \int dt_2 \tilde{F}(t_1,t_2)\Theta(t_1-t_2)\right]\, ,\\
   \tilde{F} &=&\left(\sin^2\xi\, \mu(t_1)\mu(t_2)\langle \tilde{x}(t_1)\tilde{x}(t_2)\rangle\right.\nonumber\\
           & & +\cos\xi\sin\xi\,\mu(t_1)\lambda(t_2)\langle \tilde{x}(t_1)\tilde{p}(t_2)\rangle\nonumber\\
     & & +\cos\xi\sin\xi\,\lambda(t_1)\mu(t_2)\langle \tilde{p}(t_1)\tilde{x}(t_2)\rangle\nonumber\\
   & & \left.   +\cos^2\xi\,\lambda(t_1)\lambda(t_2)\langle \tilde{p}(t_1)\tilde{p}(t_2)\rangle 
\right)\, .\nonumber
 \end{eqnarray}
 We can apply eq. (\ref{eq_apA_Cchfromgenerator}) to this form of the generating function and get
 \begin{eqnarray}
  C_{\rm ch}(t)&=&\sin^2\xi\left[  \langle x(t)x(0) \rangle
                   -\cos^2\xi\langle p(t)x(0) \rangle^2 \right]\\
           & &\times \langle e^{ip(t)} e^{-ip(0)} \rangle \, .\nonumber
 \end{eqnarray}

\end{document}